\documentclass[floatfix,secnumarabic,amssymb,nobibnotes,nofootinbib,preprint,aps,pra,showpacs]{revtex4-1}
\usepackage{epsfig}
\usepackage[colorlinks=true, citecolor=blue, linkcolor=blue, urlcolor=black]{hyperref}
\usepackage{cleveref}
\usepackage{times}
\usepackage{mathrsfs}
\usepackage{graphicx}
\usepackage{dcolumn}
\usepackage{color}
\usepackage{bm}
\usepackage{graphics,psfrag}
\usepackage{graphicx,psfrag}
\usepackage{amsmath}
\usepackage{amssymb}
\usepackage{caption}
\usepackage{subcaption}
\usepackage{soul}

\usepackage{lineno,hyperref}

\newcommand{\be}{\begin{equation}}
\newcommand{\ee}{\end{equation}}
\newcommand{\bea}{\begin{eqnarray}}
\newcommand{\eea}{\end{eqnarray}}
\newcommand{\ba}[1]{\begin{array}{#1}}
\newcommand{\ea}{\end{array}}

\usepackage{braket}

\begin{document}

\title{Exploring the route from leaky Berreman modes to bound states in continuum}
\author{Ghanasyam Remesh$^{1}$, Pravin Vaity$^{2}$, Venu Gopal Achanta$^{2}$ and Subhasish Dutta Gupta $^{1,3}$ \footnote{Corresponding author, e-mail: sdghyderabad@gmail.com}}
\affiliation{$^{1}$School of Physics, University of Hyderabad, Hyderabad-500046, India.
	\\$^{2}$Department of Condensed Matter Physics and Material Science, Tata Institute of Fundamental Research, Mumbai 400005, India.
	\\$^{3}$Tata Centre for Interdisciplinary Sciences, TIFRH, Hyderabad 500107, India.
}
\def\zbf#1{{\bf {#1}}}
\def\bfm#1{\mbox{\boldmath $#1$}}
\def\hf{\frac{1}{2}}

\begin{abstract}
             We study coupling of leaky Berreman modes in polar dielectric films (SiO$_2$) through a thin metallic  layer (gold) and show the familiar signatures of normal mode splitting. Due to very large negative real part of the dielectric function of gold, the splitting shows up only for extremely thin coupling layers. In contrast, coupling of Berreman modes through a dielectric spacer layer reveals novel possibilities of having bound states in continuum, albeit in the limit of vanishing losses. It is shown that the corresponding dispersion branches of the symmetric and antisymmetric modes can cross. BIC is shown to occur on one of these branches which is characterized by lower loss. In fact the BIC corresponds to the point where the radiative losses are minimized. For thicker  layers (both spacer and the polar dielectric) BIC is shown to occur on the higher order dispersion branches. The origin of BIC is traced to the Fabry-Perot type mechanism due of the excitation of the leaky guided modes in the central layer.   
\end{abstract}

\maketitle
\section{Introduction}

In recent years there has been a great deal of interest in the bound states in continuum (BIC) \cite{DouglasStone2016,Koshelev:20,azzam2021,Monticone2018,Li2016,Sadreev2020,Sadreev2018}. The interest in BICs is motivated by the complete confinement of the associated fields leading to small mode volume with no leakage, resulting in extremely large quality factor and local field enhancements. Recall that all these features facilitate efficient light-matter interaction. Moreover, these discrete bound states, also referred to as embedded eigenstates, coexist with the continuum. It is not surprising that the generic nature of BIC's irrespective of the specialized area have found many practical applications. Recent reviews  \cite{DouglasStone2016,azzam2021} summarize some of the areas where BIC's have been discovered starting from the pioneering works of von Neumann and Wigner, Stillinger and Herrick, Friedrich and Wintgen \cite{vonNeuman1929,Stillinger1975,Friedrich1985}. Perhaps the simplest possible structure that can lead to BIC exploits the Fabry-Perot (FP) type systems supporting coupled resonances \cite{Borisov2008,Fan2003}. This could be a generic symmetric coupled waveguide/plasmonic system. Ocurrence of BIC in such structures is closely linked to the epsilon near zero (ENZ) behavior of the films that support them. The literature on the ENZ materials and their various posibilities and applications is truly vast \cite{Shalaev2019,Engheta2017,Reshe2019,Alu2006}. BIC has been reported in a single ENZ slab \cite{Li2017} and also in spherical core-shell  or layered spherical structures with ENZ constituents \cite{silveirinha2014, monticone2014}. Very recent studies report the possibility of multiple BICs in two identical ENZ films separated by a dielectric \cite{Li2016, sakotic2020}. The ENZ behavior was modeled with a Drude model ignoring the intrinsic losses.  Since the BIC can be considered as a limiting case with zero leakage and null line width with diverging quality factor, it is pertinent to investigate the route from standard leaky modes (resonances with finite width) to the BIC (resonance with null width). 
\par
In this context we pick a polar dielectric film on a metal or dielectric substrate, which  can support the Berreman modes. Since the pioneering work of Berreman  \cite{Berreman1963}, these modes have been studied in detail both theoretically and experimentally \cite{Shaykhutdinov2017,Vassant2012a, Harbecke1985,Bichri1997,Bichri1993,Newman2015}. Several studies have been carried out to understand the non-standard properties of the leaky modes \cite{Marcuvitz1956,Alu2015,Hu2009}. Existence of the Berreman modes is also connected to the epsilon near zero (ENZ) behavior. There are several reports of Berreman modes in ENZ films \cite{Vassant2012a,Passler2019,Chen2013,Caligiuri}. A detailed analysis of the dispersion and the field profiles of these modes in an SiO$_2$ film on a gold substrate has been presented by Vassant \textit{et al.} \cite{Vassant2012a}. Thereafter, other studies highlight the local field enhancement and associated applications in sensing and nonlinear optics \cite{Passler2019,Bello2017}. Only very recently the Berreman modes have been linked to the ocurrence of BIC in the context of a Drude model \cite{sakotic2020}. In the same reference a polar dielectric (SiC) was considered briefly to show the feasibility of high-Q quasi-BIC in presence of losses. In this paper we study coupled Berreman modes where the coupling is mediated by (a) a metal (Au) film and (b) a dielectric (air) layer. It is clear that   (a) leads to an evanescent coupling with usual normal mode splittings and avoided crossings as with coupled plasmons of a thin metallic film \cite{sdgbook}. Our major interest is in scheme (b) where the polar dielectric films near the epsilon near zero (ENZ) condition act like ideal mirrors of the FP cavity, leading thereby to the possibility of BIC \cite{Li2016,sakotic2020}. Our main target is to explore the route from the leaky Berreman modes to the BIC. In particular, in order to establish the connection of the leaky symmetric or the antisymmetric mode (hereafter symmetry will relate to the symmetry of the magnetic field distribution) to the BIC we ignore the intrinsic losses in the SiO$_2$ film. We show that, depending on the symmetry of the mode (symmetric or antisymmetric) a continuous tuning of the operating point moves the leaky mode to the BIC. In other words the BIC can be recognised as the limiting case of the one of the eigen modes which can exhibit null radiation leakage. The results obtained are explained by a generic Hamiltonian as in \cite{Haus1983}. We also studied the leaky guided modes of the spacer layer with larger thickness to show analogous effects with the symmetric and antisymmetric branches of the dispersion. The results are also validated by direct calculation of the reflectivity and superposing the dispersion data on it.
\par 
The structure of the paper is as follows. In Section \ref{sec:level2}, we present the formulation of the problem. Section \ref{sec:level3} presents the numerical results for both the schemes (a) and (b) and discusses the results. In Conclusion we summarize the main findings of the paper.	
\section{\label{sec:level2}Formulation of the problem}	
\begin{figure}
	\centering
	\includegraphics[width=0.8\linewidth]{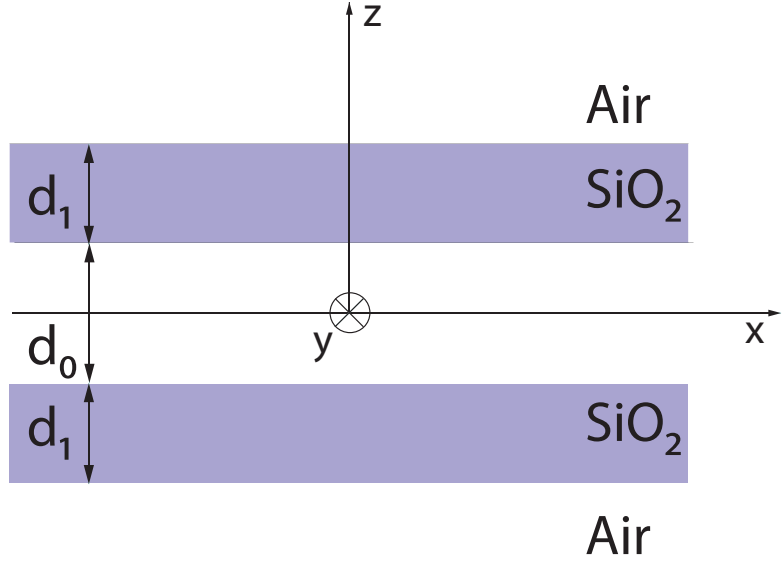}
	\caption{Schematics of the symmetric structure with Silica-Metal/Dielectric-Silica layers. The  medium outside is assumed to be air.}
	\label{fig1}
\end{figure}

Consider the symmetric structure shown in Fig.  \ref{fig1} where two thin films of polar dielectric (e.g. SiO$_2$) are separated by a spacer layer of metal (Au) or dielectric (e.g. air). Let the dielectric function of Au and SiO$_2$ be given by \cite{Vassant2012a,Meyer2006}
\begin{equation}
\label{eq1}
\begin{split}
\epsilon_{Au} (\lambda) &= \epsilon_{Au,\infty} - \frac{1}{\lambda_p^2(1/\lambda^2 
	+ i/\gamma_p\lambda)}\\
& +  \sum_{i =1,2} \frac{A_i}{\lambda_i}\left[  \frac{e^{i\Phi_i}}{
	(1/\lambda_i-1/\lambda - i/\gamma_i)}  +  \frac{e^{-i\Phi_i}}{
	(1/\lambda_i+1/\lambda + i/\gamma_i)} \right],
\end{split}
\end{equation}
where $\epsilon_{Au,\infty} = 1.53$, $\lambda_p = 0.145 \mu$m , $\gamma_p = 17.0\mu$m, $A_1 = 0.94$,  $\phi_1 = -\pi/4$, $\lambda_1 = 0.468\mu$m, $\gamma_1 = 2.3\mu$m , $A_2 = 1.36$,  $\phi_2 = -\pi/4$, $\lambda_2 = 0.331\mu$m, $\gamma_2 = 0.94 \mu$m, and
\begin{equation}
\label{eq2}
\epsilon_{SiO2} = \epsilon_{\infty} \frac{\omega^2 - \omega_L^2 + i\omega \Gamma}{\omega^2 - \omega_T^2 + i \omega \Gamma } ,
\end{equation}
where $\epsilon_{\infty}  =  2. 095$, $\omega_L = 2.30\times 10^{14} $ rad/s ($\lambda_L=8.19 \,\mu m$), $\omega_T =  1.98\times 10^{14}$ rad/s ($\lambda_T = 9.51\,\mu m$) and $\Gamma = 1.35\times 10^{13}$ rad/s ($\lambda_\tau =139.53\,\mu m$). 
\par
We have carefully studied the applicability of models given by Eqs.(\ref{eq1}) and (\ref{eq2}) over the range of frequencies in our study. Eq.(\ref{eq1}) for gold was designed to fit the experimental data of Johnson and Christie \cite{Johnson1972}, which does not cover our frequency range. However, a later study by Babar and Weaver \cite{Weaver2015} covers a much wider range including our spectral domain. A comparison of the results for the real and imaginary parts of the dielectric function of gold from these experimental sources with the model in their respective ranges of wavelength reveals that over the range of wavelengths considered in \cite{Johnson1972} the fit is truly good, while in the extended range of \cite{Weaver2015} it is qualitative retaining the essential features of gold at larger wavelengths (not shown). As will be shown below in section \ref{sec:level3.1}, the important feature is the large and negative value of the real part of dielectric function of gold, which is adequately captured by Eq.(\ref{eq1}), yielding the correct physics.
\par
\begin{figure}[h]
	\includegraphics[width=0.74 \linewidth]{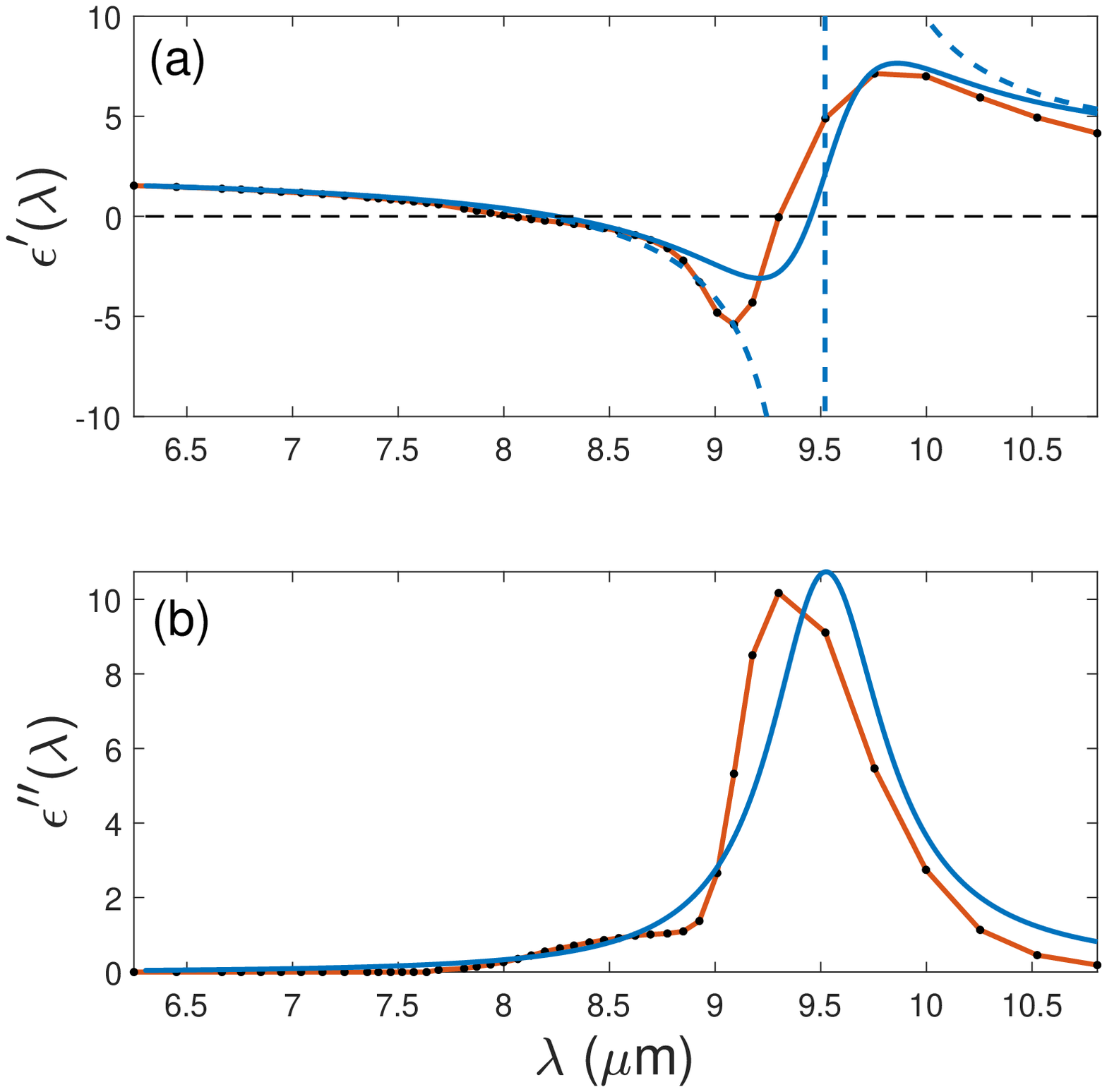}
	\centering
	\caption{(a) Real and (b) imaginary parts of the permittivity of SiO$_2$ as functions of wavelength. Blue solid line is from Eq.(2), and the red line with dots are the experimental data taken from \cite{Philipp1985}. The ENZ behavior due to $\omega=\omega_L$ can be seen near $\lambda\sim8.12\,\mu m$. The resonant behavior at $\omega_T$ is near $\lambda\sim9.52\,\mu m$. The dashed blue line is the value of $\epsilon_{SiO_2}$ without damping ($\Gamma=0$).}
	\label{Fig:1.5}
\end{figure}

 
The model for SiO$_2$ (Eq.(\ref{eq2})) with the parameters mentioned in the text was checked against the experimental data of Philipp \cite{Philipp1985} and as can be seen from the comparison in Fig. \ref{Fig:1.5}, is an acceptable model. What is important is that the ENZ behavior (Real($\epsilon_{SiO_2})=0$ near $\lambda\sim 8.12 \,\mu$m corresponding to $\omega=\omega_L$) occurs nearly at the same location in Fig. \ref{Fig:1.5}. Recall that for BIC we need to look at a lossless system with $\Gamma=0$ in Eq.(\ref{eq2}). The corresponding $\epsilon_{SiO_2}$ is shown by the dashed line in Fig. \ref{Fig:1.5}(a).     
\par
The generic features of a polar dielectric (like SiO$_2$) given by  Eq.(\ref{eq2}) show that close to the  frequency $\omega_L$, one has an epsilon near zero behavior, while a resonant behavior is exhibited at $\omega = \omega_T$ (see Fig. \ref{eq2}(a)). The ENZ behavior can lead to interesting electromagnetic phenomena, which can be easily seen if one considers an interface between a conventional dielectric (say, with $\epsilon_1$) and a polar dielectric (with $\epsilon_2$). Due to the boundary conditions for the normal component of the electric field, one can expect a large enhancement of the field ($E_2^{\perp} = \epsilon_1/\epsilon_2E_1^\perp$ as $\epsilon_2$ tends to zero). Further, as shown in \cite{Vassant2012a}, one can expect the presence of two modes (inside and outside the lightcone) near the ENZ condition $\epsilon_{\text{SiO}_2}(\omega) = 0$. The first mode is referred to as the Berreman mode (which is leaky in nature), and it is responsible for the large absorption in thin polar dielectric films first demonstrated by Berreman in his seminal paper \cite{Berreman1963}. The other mode (evanescent in character), present outside the lightcone, was labeled as the ENZ mode due to its proximity to the $\omega=\omega_L$ line. It was shown that the ENZ mode was related to the long-range surface modes, and exhibits a nearly dispersionless behavior for thin films. It is clear from the evanescent nature of this mode that one cannot couple incident plane waves to the evanescent ENZ mode via oblique incidence due to the momentum mismatch, unless one uses a high index prism or a grating.
\par 
We use the analytical expressions of the dielectric functions of both metal  and polar dielectric so as to enable us to calculate the roots of the dispersion relations in order to make connection with the existing results and validate our codes. We first consider the manifestations of the coupling through the gold film retaining the full losses in the SiO$_2$ layer. We focus only on the coupled Berreman modes. As will be shown below, the coupling leads to the familiar normal mode splitting.

\par
It is pertinent to mention here that we used a standard characteristic matrix method \cite{sdgbook} to calculate the reflection/transmission characteristics of the structure for a p-polarized plane wave incidence at an angle $\theta_i$. The corresponding symmetric and antisymmetric dispersion relation for the structure (shown in Fig. \ref{fig1}) is given by \cite{sdgbook}
\begin{eqnarray}
\label{eq3}
D_{sym}(k_x, \omega)&=&m_{21}+m_{22}p_{tz}=0\\\label{eq4}
D_{antisym}(k_x, \omega)&=&m_{11}+m_{12}p_{tz}=0
\end{eqnarray}
where $p_{jz}= {k_{jz}}/{k_0\epsilon_{rj}}$ is the normalized transverse wave vector, given by $k_{jz} = \pm\sqrt{\epsilon_{rj}k_0^2-k_x^2}$, and $\epsilon_{rj}$ is the relative dielectric permittivity of the $j^{th}$ medium. The correct branch of $k_{jz}$ is chosen such that $\text{real}(k_{jz})+\text{imag}(k_{jz})>0$ \cite{Vassant2012a}.  $m_{ij}$ are the elements of the characteristic matrix for the upper half of the structure $M_{T}=M_0(d_0/2)M_1(d_1)$, with $M_j$ being the characteristic matrix of the $j$th layer.
In Eqs.(\ref{eq3}) and  (\ref{eq4}) $k_x$ is the surface component of the wave vector which is assumed to be real as in \cite{Vassant2012a}. We investigated the roots (complex $\omega$) of the dispersion relation by using standard root finding algorithm. The results were validated by additional Lumerical FDTD simulations of the same structure.
\par
As mentioned in the Introduction our aim is to establish a link between the Berreman modes and possible BIC's in such structures. Keeping in view of the fact that pure BIC in quantum systems occurs in absence of any losses, we look at the coupled system with dielectric spacer layer in the limit of vanishing material losses ($\Gamma=0$). We show below that the BIC occurs on one of the branches of the coupled system. Note that BIC in analogous systems of two coupled metal films with null losses has been reported using ENZ behavior of metals at the plasma frequency \cite{Li2016,sakotic2020}. The details of our investigation and the results are presented in the next section.
\section{\label{sec:level3}Numerical results and discussions}
In order to validate our code we first recovered the results presented by Vassant \textit{et al.} \cite{Vassant2012a} for a silica film on a gold substrate (not shown). Since our target is to understand the link between the leaky modes and BIC, we focus our attention only to the Berreman modes occurring  on the left of the light line ($k_x<\sqrt{\epsilon_i
}k_0$) close to the ENZ condition $\omega\sim\omega_L$. The results for the middle gold and air film are presented separately in the following subsections.
\subsection{\label{sec:level3.1}Coupling via evanescent waves}
\begin{figure}
	\centering
	\includegraphics[width=0.8\linewidth]{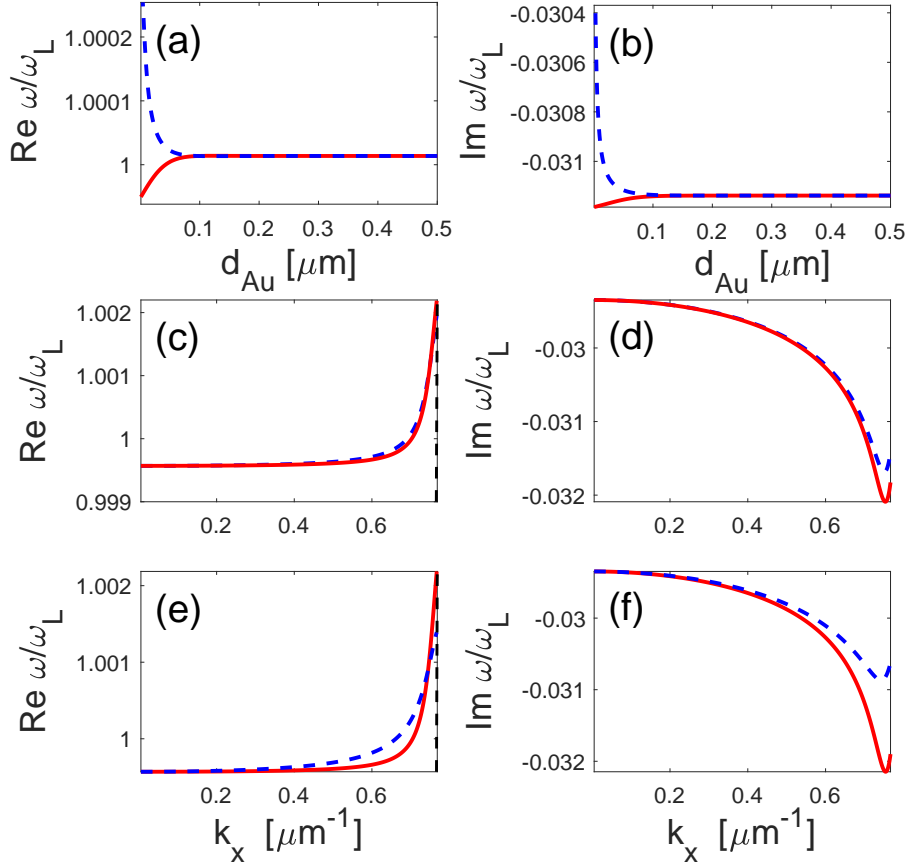}
	\caption{(a) Real and (b) imaginary parts of the normalized frequency $\omega/\omega_L$ for symmetric (solid red) and antisymmetric (dashed blue) modes as functions of the gold thickness, $d_0=d_{\text{Au}}$ for fixed surface component of wave vector $k_x=0.7~\mu m^{-1}$. (c) Real and (d) imaginary part of $\omega/\omega_L$ as functions of of $k_x$ for $d_{1}=d_{\text{SiO}_2}=0.02~\mu m$ and $d_{0}=0.02~\mu m$.	(e),(f) same as in (c),(d) except for $d_{0}=0.005~\mu m$. The black dashed line represent the lightline $k_x = \displaystyle \omega/c$.} 
	\label{Fig:2}
\end{figure}

It is well known that at lower frequencies ($\omega<< \omega_p$), the real part of the dielectric function of gold is large and negative indicating a very small skin depth. Thus coupling of the Berreman modes is possible only for extremely low thickness of the coupling layer. The coupling manifests itself as normal mode splittings into the symmetric and the antisymmetric modes. As mentioned earlier symmetry will be defined by the symmetry of the $p-$polarized  magnetic fields. The results for the normal mode splittings for the real and imaginary parts of the roots of the dispersion relations Eqs.(\ref{eq3}) and (\ref{eq4}) are shown in Fig. \ref{Fig:2}. We have plotted the real (imaginary) parts of the roots  in the left (right) column. The top row shows these roots as functions of $d_0$ (Figs. \ref{Fig:2}(a),(b)) for fixed $k_x=0.7~\mu m^{-1}$. The middle (bottom) row depicts the same  as functions $k_x$ for fixed $d_0=0.02~\mu m$ in Figs. \ref{Fig:2}(c),(d) ($d_0=0.005~\mu m$ in Figs. \ref{Fig:2}(e),(f)). The results are analogous to surface plasmon coupling  in a thin metal film. For the parameters of SiO$_2$ and for large gold thickness $\sim 15$ nm, there was no experimental evidence of coupling between the Berreman modes \cite{Bichri1997}. The splitting shows up for gold film thickness of about $d_0 \lesssim 0.005~ \mu m$.     As can be seen from these figures, the coupling induced splittings are more prominent for lower  thickness of the gold film, since the thin metal film allows for better coupling between the modes. These features are all expected, though they were not well studied in the context of Berreman modes. As will be shown below, novel features appear when the spacer layer between the two polar dielectric films is a dielectric. In order to avoid complications due to material dispersion, we assumed the spacer layer to be air and the communication between the Berreman modes is established by propagating waves. 
\subsection{\label{sec:level3.2}Coupling via propagating waves}
The results of the previous subsection were obtained retaining the full losses both in SiO$_2$ and the Au layers. The pure BIC can exist only in systems without losses. Keeping this in view we consider a theoretical model where we study the same structure albeit with vanishing losses ($\Gamma = 0$). In short, we study Fabry-Perot type BIC mediated by the modes of the symmetric structure where two thin SiO$_2$ films with vanishing losses are separated by an air gap. The coupling via the propagating waves leads to the symmetric and the antisymmetric modes. These modes are calculated exploiting the symmetry of the structure as in \cite{sdgbook}. 
\begin{figure}
	\centering
	\includegraphics[width=0.8\linewidth]{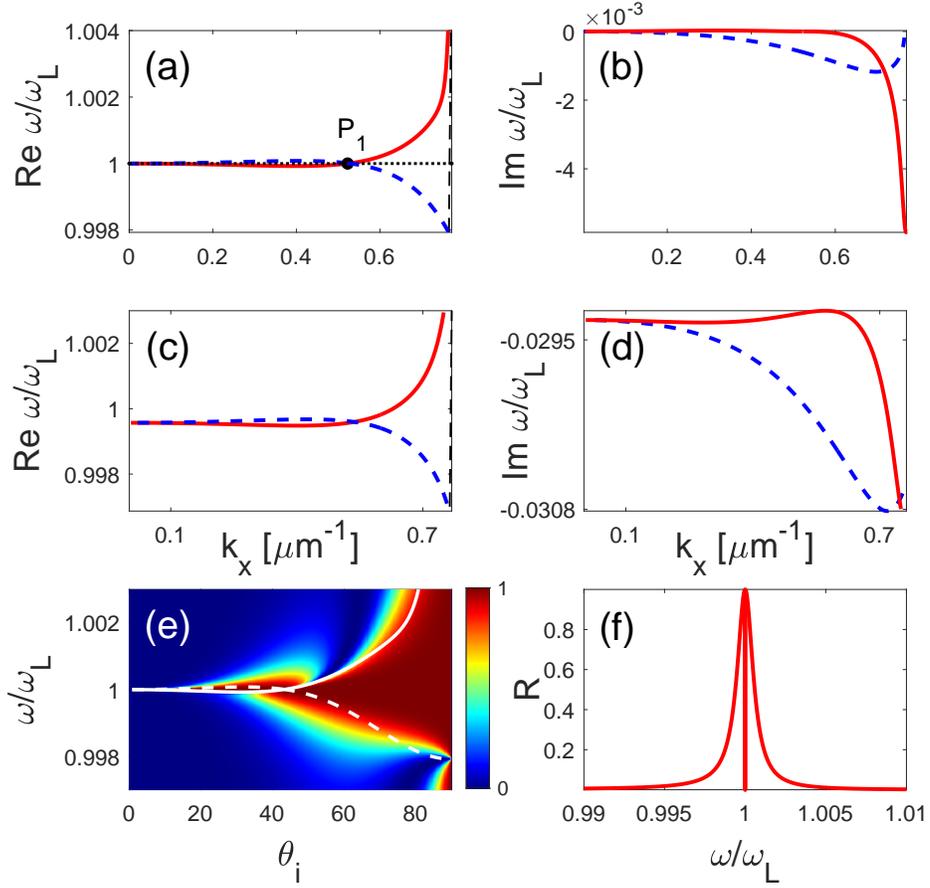}
	\caption{(a) Real and (b) imaginary parts of the normalized frequency $\omega/\omega_L$ as functions of $k_x$ for $d_1=d_{\text{SiO}_2}=0.02~\mu m$ and $d_0=d_{\text{Air}} = 5.6 ~\mu m$. The BIC is marked by a black point P$_1$ in (a). (c),(d) Same as in (a),(b), but retaining full losses in the SiO$_2$ layers.	The red solid (blue dashed) line depicts the symmetric (antisymmetric) mode. The black dashed (dotted) line represents the lightline (ENZ line  $\omega/\omega_L=1$). (e) Shows the intensity reflection $R$ as a function of angle of incidence and frequency. The dispersion curves for symmetric (white solid) and antisymmetric (dashed white) are superimposed. (f) Intensity reflection as a function of the frequency at a fixed angle of incidence ($\theta_i\simeq 42.96^0$ at which BIC occurs).}.
	\label{Fig:3}
\end{figure}
\par
We now pay attention to the central result of this study, namely, the occurrence of the BIC on either of the dispersion branches having the least radiation leakage. For very thin SiO$_2$ layers the two dispersion branches for the real part can cross. Crossing in the imaginary part albeit at a different value of $k_x$ is also observed. The possibility of crossing of the real or imaginary parts of the dispersion curves has been discussed in detail \cite{Rotter2003}. In a different context of exceptional points in non-Hermitian systems crossing of the dispersion branches for the real part has been pointed out \cite{Alu2019}. Such crossing in the real part is associated with the repulsion of the branches for the imaginary part. Another important aspect is the overlap of the crossing of the dispersion curves with the ENZ line, Re$(\omega)=\omega_L$ in the $\omega$--$k_x$ plane. Note that when the ENZ condition is met, the polar dielectric layers act as ideal mirrors with unity reflectivity, thus meeting the requirement for the Fabry-Perot type BIC.  The results for the real and imaginary parts of the roots of the dispersion relation for complex $\omega$ (normalized to $\omega_L$) for $d_0= 5.6~\mu m $, $d_1= 0.02 ~\mu m$ are shown in Figs. \ref{Fig:3}(a) and \ref{Fig:3}(b), respectively. Recall that the presence of the finite imaginary part of the root implies losses of the modes in a system where intrinsic material losses were ignored. Here losses owe their origin to the leaky character of the modes. It is interesting to note that for Berreman modes of very thin films, a fine tuning of the system parameters can make the crossing occurs on the ENZ line. The BIC is then realized on either of the symmetric or the antisymmetric branch, depending on which mode can exhibit the least leakage (null loss). It is clear from Fig. \ref{Fig:3}(a) that for the said system parameters, the BIC occurs on the symmetric mode at $k_x\simeq 0.5225 ~\mu m^{-1}$ having null leakage (see Fig. \ref{Fig:3}(b). This is validated by direct calculation of the reflectivity (for incidence of a plane p-polarized light at an angle $\theta_i$) as a function of frequency and angle of incidence and superposing the real part of the dispersion curves on it. While superposing the real root on the reflectivity data, we have moved from $k_x$ to angle of incidence $\theta_i$ for better clarity. BIC can be noted at the crossing of the white solid and dashed lines (dispersion data), where the blue region shrinks to null width  (see Fig. \ref{Fig:3}(e)).  Fig. \ref{Fig:3}(f) shows the intensity reflection profile for a fixed angle of incidence $\theta_i\simeq42.96^0$.  It is thus clear that for the said system parameters the symmetric Berreman mode has vanishing leakage  at $k_x\simeq 0.5225 ~\mu m^{-1}$, and it is the one which has evolved to BIC coexisting with the leaky antisymmetric mode. Note that because of the crossing of the dispersion branches, the same value of $k_x$, namely $k_x\simeq 0.5225 ~\mu m^{-1}$,  corresponds to the momentum matching for the excitation of the leaky antisymmetric mode.
\par
\begin{figure}
	\centering
	\includegraphics[width=0.8\linewidth]{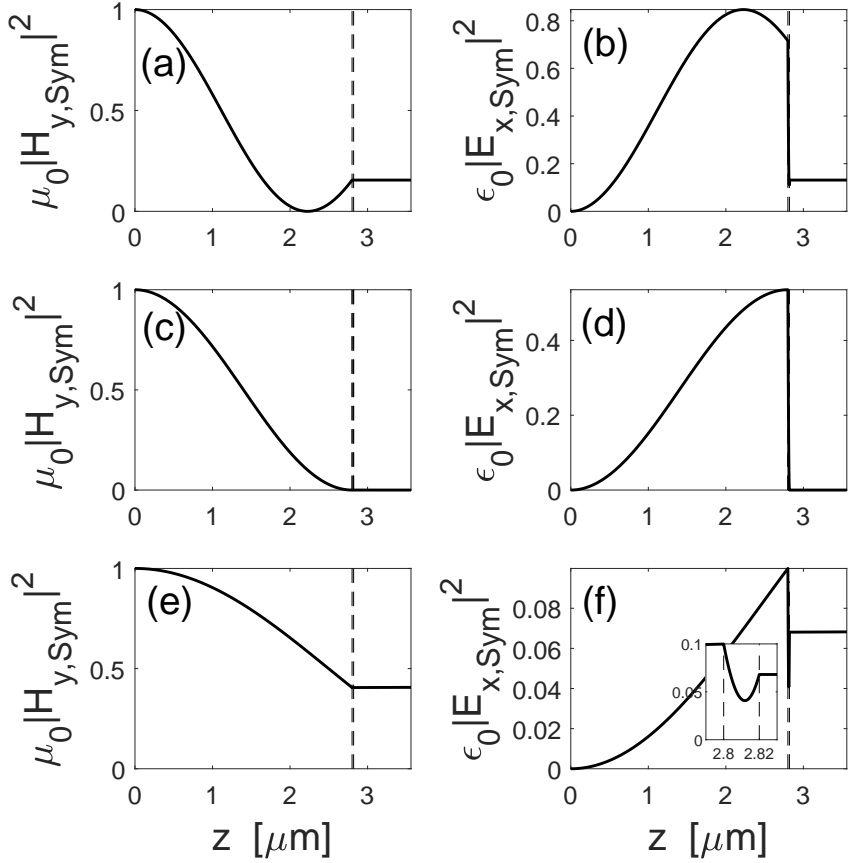}
	\caption{Spatial distribution of the tangential field intensities (a) $\mu_0|H_y|^2$ and (b) $\epsilon_0|E_x|^2$ of the symmetric eigenmodes for the structure studied in Fig. \ref{Fig:3}. (a),(b) before BIC (at $k_x=0.3~\mu m^{-1}$).   (c),(d) at BIC (P$_1$ at $k_x\simeq 0.5225~\mu m^{-1}$) and (e),(f) after BIC (at $k_x=0.7~\mu m^{-1}$). The dashed vertical lines represent the boundaries between two different layers. The inset in (f) shows an expanded part near the boundary to reflect the continuity of the fields. The other parameters are as in Fig. \ref{Fig:3}.}
	\label{Fig:4}
\end{figure}

 
In order to confirm the total confinement of the fields at BIC, we have plotted the eigen mode spatial profiles at and away from the BIC. The results for upper half of the structure are shown in Fig. \ref{Fig:4} for the tangential intensity profiles $\mu_0|H_y|^2$ and $\epsilon_0|E_x|^2$ as functions of distance from the center. The inset in Fig. \ref{Fig:4}(f) shows an expanded portion just to verify the continuity of the tangential component of  $|E_x|$. It is clear that away from BIC one excites the leaky modes while the BIC corresponds to total confinement with null leakage.
\par 
In order to understand what happens in a realistic system, we repeated the above calculations retaining the losses ($\Gamma = 1.35\times 10^{13} ~rad/s$). The results are shown in Figs \ref{Fig:3}(c) and \ref{Fig:3}(d). A comparison of Fig. \ref{Fig:3}(c) with Fig. \ref{Fig:3}(a) reveals that, despite the persistence of the crossing on the ENZ line both the split modes have losses ruling out the possibility of BIC or a quasi-BIC with high Q-factor. In fact, the highest quality factor for the symmetric Berreman mode for a realistic SiO$_2$ film of thickness $0.02$ $\mu m$ with a gap of $5.6\,\mu m$ between them  was estimated to be $\sim 17$. Moreover, the minimum of the absorption for the symmetric mode occurs away from the crossing point. This is also verified by direct reflectivity calculations (not shown). These results, along with lack of experimental evidence of normal mode splittings may suggest that SiO$_2$ may not be the ideal material for practical applications due to its large intrinsic losses. However, for other polar materials like AIN, a quality factor of $\sim 81$ was reported experimentally (see Fig. 2 in \cite{Passler2019}). The feasibility of quasi-BIC with quality factors as high as $10^3$ have been reported in other structures \cite{sakotic2020,Sadrieva2017}.
\par
\begin{figure}
	\centering
	\includegraphics[width=0.8\linewidth]{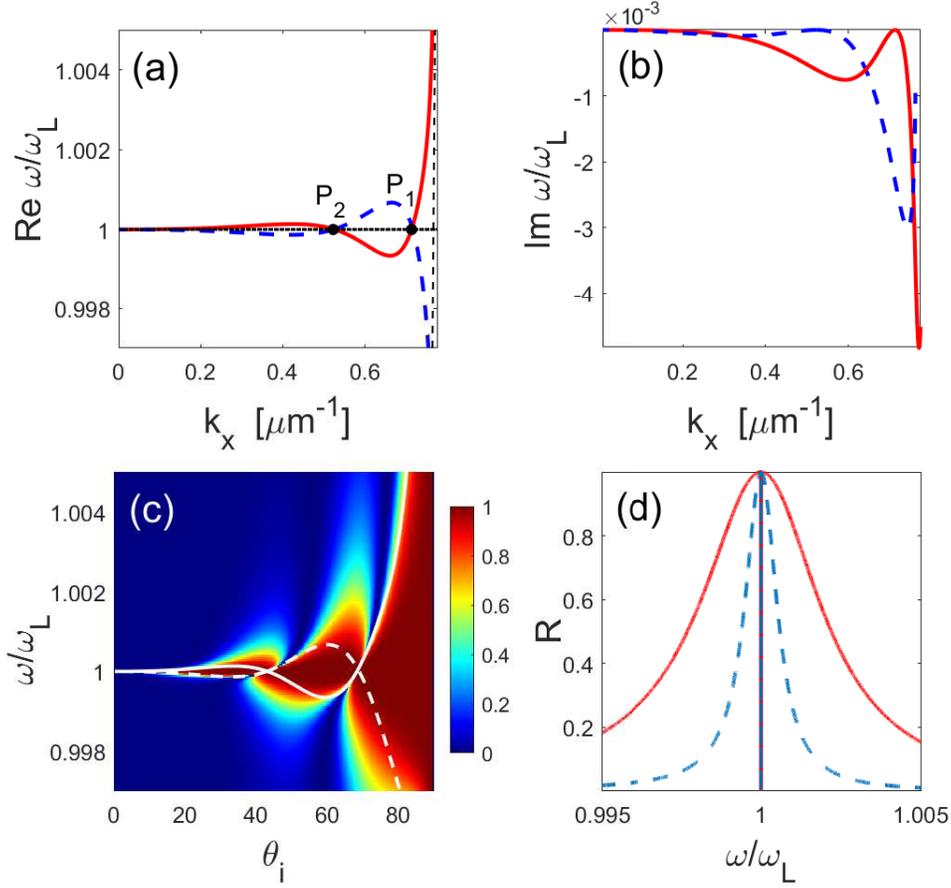}
	\caption{(a),(b),(c) The same as in Fig. \ref{Fig:3}(a),(b),(e), respectively, except for $d_0=d_{\text{Air}} = 11.2 ~\mu m$. Note the 2nd (1st) order BIC on the antisymmetric (symmetric) branch, marked by P$_2$ (P$_1$). (d) The red solid (blue dashed) line shows the intensity reflection as a function of the frequency at a fixed angle of incidence $\theta_i\simeq 68.52^0$ ($\theta_i\simeq42.96^0$) corresponding to the 1st (2nd) order BIC.}
	\label{Fig:5}
\end{figure}

\begin{figure}
	\centering
	\includegraphics[width=0.8\linewidth]{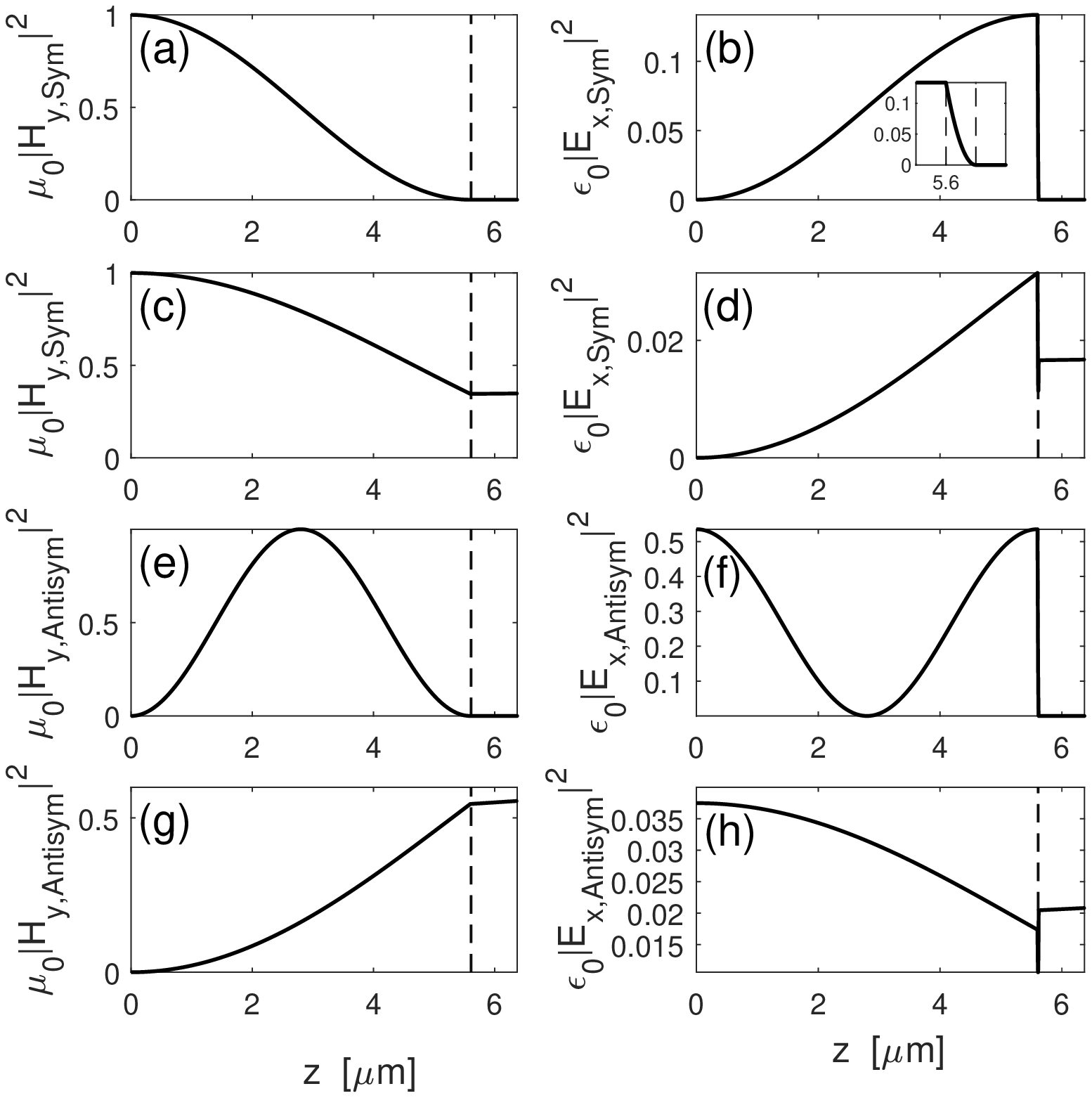}
	\caption{Spatial distribution of the tangential field intensities of (a) $\mu_0|H_y|^2$ and (b) $\epsilon_0|E_x|^2$ of the symmetric eigenmodes for the structure studied in Fig. \ref{Fig:5}(a), (a),(b) at $1^{st}$ order BIC ( P$_1$ at $k_x\simeq0.7135~\mu m^{-1}$),  (c),(d)   after the 1st order BIC (at $k_x = 0.75~\mu m^{-1}$). (e),(f) Antisymmetric eigenmode intensity distributions at $2^{nd}$ order BIC (at $k_x\simeq0.5225~\mu m^{-1}$) and (g),(h) after $2^{nd}$ order BIC (at $k_x=0.75~\mu m^{-1}$). The inset in (b) shows an expanded part near the boundary to reflect the continuity of the fields.The other parameters are as in Fig. \ref{Fig:5}.}
	\label{Fig:6}
\end{figure}

 
We now show that it is possible to have the BIC occurring on both the antisymmetric and symmetric branches. To this end we double the central air layer thickness accomodating the possibility of higher order symmetric and antisymmetric modes. In particular we take the following system parameters: $d_1=d_{\text{SiO}_2} = 0.02 ~\mu m
$, $d_0=d_{\text{Air}} = 11.2~\mu m$. The results are shown in Fig. \ref{Fig:5}(a) and \ref{Fig:5}(b). The corresponding reflectivity spectra and the reflectivity profiles are shown in Figs. \ref{Fig:5}(c) and \ref{Fig:5}(d), respectively. As can be seen from these figures that now the BIC occurs at two points. While the first occurs on the antisymmetric Berreman branch at $k_x\simeq0.5225~\mu m^{-1}$, the second shows up on the symmetric branch at values of $k_x\simeq 0.7135 ~\mu m^{-1}$ corresponding to vanishing leakage for these modes. Note also that the symmetric Berreman branch can not lead to BIC at the first crossing at $k_x\simeq0.5225 ~\mu m^{-1}$ due finite leakage. The remaining features are the same as in Fig. \ref{Fig:3}. For completeness we have presented the eigen mode spatial profiles at and away from BIC in Fig.  \ref{Fig:6}.
\par
\begin{figure}
	\centering
	\includegraphics[width=0.8\linewidth]{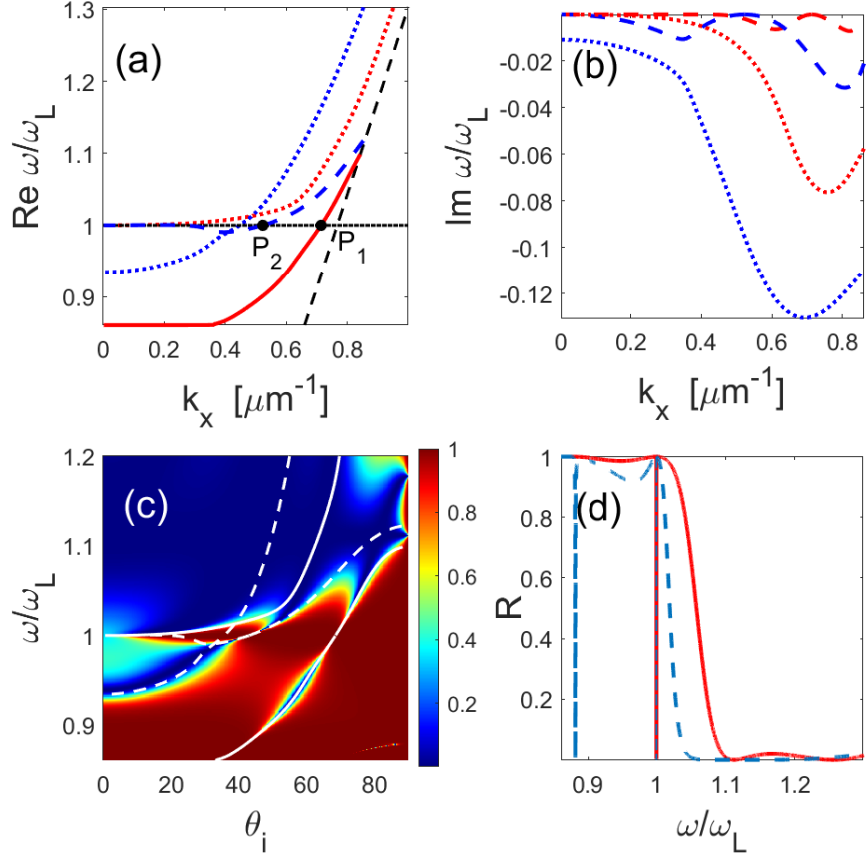}
	\caption{(a),(b),(c) The same as in Fig. \ref{Fig:5}(a),(b),(c), but for $d_{\text{SiO}_2} = 1.0 ~\mu m$ and $d_{\text{Air}} = 11.2 ~\mu m$. The dotted red (blue) line represents the coupled symmetric (antisymmetric) Berreman mode. The red solid (blue dashed) line represents symmetric (antisymmetric) higher order leaky guided mode. The 2nd (1st) order BIC lies on the antisymmetric (symmetric) branch, marked by P$_2$ (P$_1$). (d)  The red solid (blue dashed) lines gives the intensity reflection as a function of frequency at a fixed angle of incidence $\theta_i \simeq 68.52^0$ ($\theta_i\simeq42.96^0$) corresponding to the 1st (2nd) order BIC.}
	\label{Fig:7}
\end{figure}

\begin{figure}
	\centering
	\includegraphics[width=0.8\linewidth]{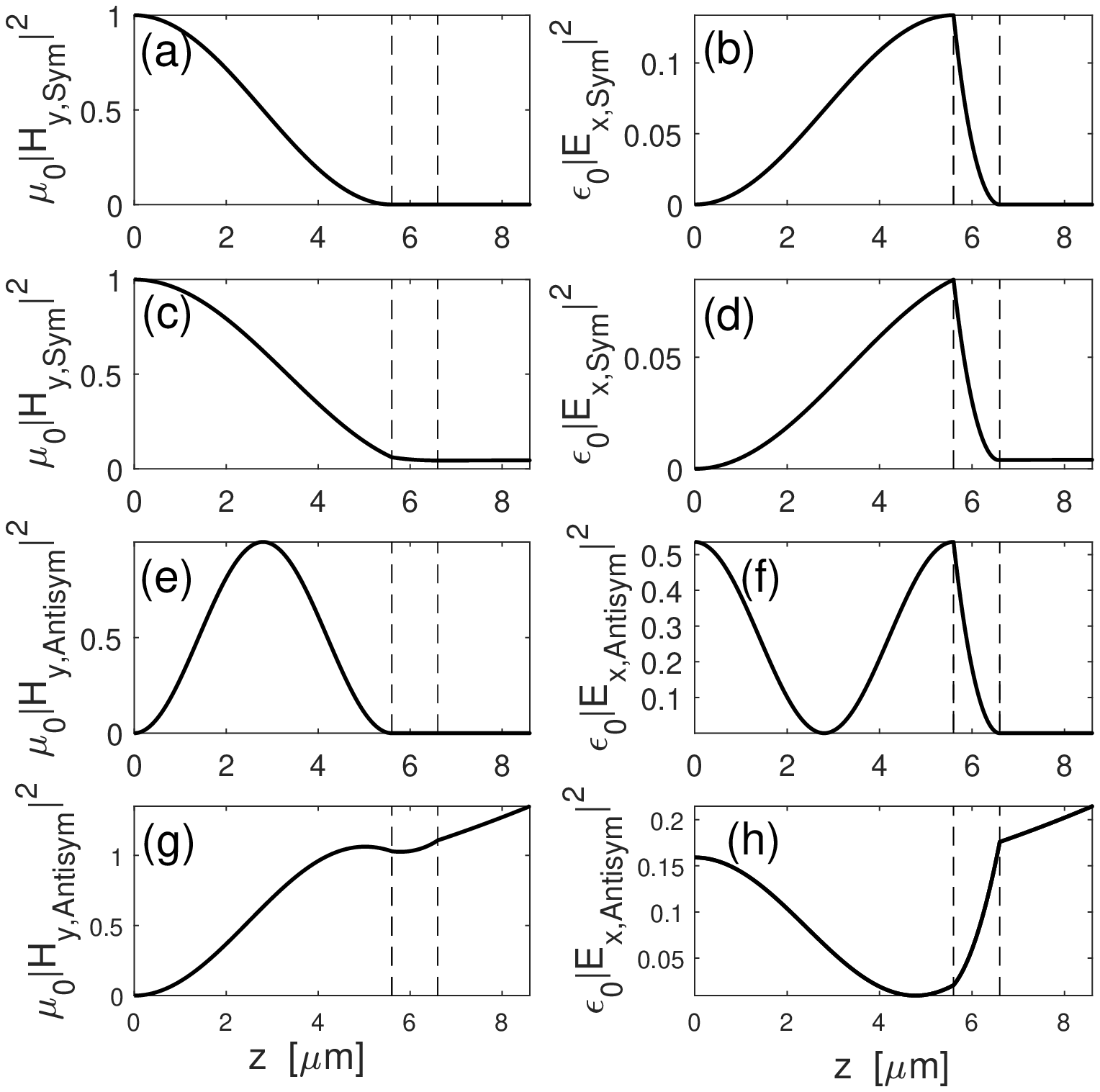}
	\caption{Spatial distribution of the tangential field intensities of the eigen modes (a) $\mu_0|H_y|^2$ and (b) $\epsilon_0|E_x|^2$ of the symmetric eigenmodes for the structure studied in Fig. \ref{Fig:7}(a). (a),(b) at $1^{st}$ order BIC (P$_1$ at $k_x\simeq0.7135~\mu m^{-1}$),  (c),(d)    after the 1st order BIC (at $k_x=0.75~\mu m^{-1}$). (e),(f) Antisymmetric eigenmode intensity distributions at $2^{nd}$ order BIC (P$_2$ at $k_x\simeq0.5225~\mu m^{-1}$)  and   (g),(h)    after $2^{nd}$ order BIC (at $k_x=0.7~\mu m^{-1}$). The other parameters are as in Fig. \ref{Fig:7}. }
	\label{Fig:8}
\end{figure}

 
The possibility of having BIC with both symmetric and antisymmetric Berreman modes for thin SiO$_2$ layers is due the fact that  both the dispersion branches oscillate near the ENZ line and  they can cross on the same. The situation is different for thicker SiO$_2$ layers, when the dispersion branches can move away from the ENZ line with larger radiative leakage. Nevertheless, some of these  dispersion branches can cross the ENZ line leading to BIC, though there is no crossing of the dispersion branches. The results for a system with $d_1=1~ \mu m$ with other parameters as in Fig. \ref{Fig:5} are shown in Fig. \ref{Fig:7}. The corresponding eigen function plots are shown in Fig. \ref{Fig:8} for both the BIC's in Fig. \ref{Fig:7}.
\par 
Results presented in Fig. \ref{Fig:3}, Fig. \ref{Fig:5} and Fig. \ref{Fig:7} can easily be understood in terms of a generic Hamiltonian given by the following
\cite{DouglasStone2016,Haus1983}
\begin{eqnarray}
H = \begin{pmatrix}
\omega_0  & g \\
g &  \omega_0
\end{pmatrix}
-i\gamma \begin{pmatrix}
1 & e^{ik_zd_0}\\
e^{ik_zd_0} &  1
\end{pmatrix},
\end{eqnarray}
where $\omega_0$ is the resonance frequency of the uncoupled resonators, $g$ is the coupling constant between them,  $\gamma$ is the radiation rate of the individual resonances and $k_zd_0$ is the propagation phase shift between the two resonators with $k_z$ being the component of the wave vector along the $z$-axis. The two eigenvalues of $H$ are
\begin{eqnarray}
\omega_{\pm} = \omega_0 \pm g - i\gamma (1\pm e^{ik_zd_0}).
\end{eqnarray}
Here, `+' (`-') represents the symmetric (antisymmetric) eigenmodes. 
When $k_zd_0$ is an integer multiple of $\pi$  one of the two eigenmodes becomes more lossy with twice the original decay rate, and the other eigenmode becomes a BIC with a purely real eigenfrequency. Odd (even) integer multiples of $\pi$ correspond to symmetric (antisymmetric) BICs. It is thus clear how when we change the order of the mode, BICs shift from the symmetric to the antisymmetric branch. For example for $d_0=5.6~\mu m$, $k_zd_0=\pi$ at $k_x\simeq0.5225~\mu m^{-1}$ leading to the BIC on the symmetric branch with $Im(\omega_+)=0$ (see Fig. \ref{Fig:3}). In contrast, for $d_0=11.2~\mu m$, $k_zd_0=2\pi$ at $k_x\simeq0.5225~\mu m^{-1}$, the BIC occurs on the antisymmetric branch with $Im(\omega_-)=0$. It is clear that 1st order BIC takes place again on the symmetric branch for $k_zd_0=\pi$ at $k_x\simeq0.7135~\mu m^{-1}$ (see Figs. \ref{Fig:5} and \ref{Fig:7}). 
\par
Indeed  for a valid ENZ condition $\epsilon_{SiO_2} = 0$ (or $\omega=\omega_L$) ), the solution to the dispersion relation (Eqs.(\ref{eq3}) and (\ref{eq4})) can be solved easily for the values of $k_x=k_0\sin\theta$ recovering the above results. For example, Eq.(\ref{eq3}) for the symmetric mode can be rewritten in the form
\begin{eqnarray}
&&-i\frac{p_{0z}}{p_{1z}}\cos (k_{1z}d_1)\sin\left(\frac{k_{z}d_0}{2}\right) - i  \sin(k_{1z}d_1)\cos\left(\frac{k_{z}d_0}{2}\right) \nonumber\\
&&+ \frac{p_{0z}}{p_{1z}}\cos(k_{1z}d_1)\cos\left(\frac{k_{z}d_0}{2}\right) - \frac{p_{0z}^2}{p_{1z}^2}\sin(k_{1z}d_1)\sin\left(\frac{k_{z}d_0}{2}\right) = 0.\label{eq7}
\end{eqnarray}
In Eq.(\ref{eq7}) we have replaced $k_{0z}$ by $k_{z}$. For $\epsilon_1=\epsilon_{SiO2} \rightarrow 0$, the terms with $1/p_{1z}=\frac{k_0 \epsilon_1}{\sqrt{k_0^2\epsilon_1-k_x^2}}= \frac{ \epsilon_1}{i \sin(\theta)}$ vanish, leading to the FP BIC condition
$$  k_{z}d_0={k_0d_0\cos\theta} = \pi $$
for the symmetric mode. Analogous procedure for the antisymmetric mode leads to the other condition. 
$$ k_{z}d_0= k_0 d_0\cos\theta =2\pi $$
\par  
All through this study we have not discussed the trivial BIC occurring at normal incidence, as expected in any symmetric structure. These are not interesting since we cannot control their position unlike the BIC's emerging from the Berreman or other higher order modes.
   \section{\label{sec:level4}Conclusions}
We have studied a symmetric layered structure involving two polar dielectric (SiO$_2$) films on both sides of a metallic (Au) or dielectric (air) film. The structure can support coupled Berreman modes. For evanescent coupling through a thin metallic film, the standard features of avoided crossing and normal mode mode splittings are reported. For the structure with the dielectric spacer layer, BIC's are shown to exist in the limit of vanishing intrinsic material losses. A thorough analysis of the dispersion branches of the symmetric and the antisymmetric modes revealed that the BIC's always occur on one of these dispersion branches which has the least radiative loss. It is also shown that for very thin SiO$_2$ films the symmetric and antisymmetric branches can cross. Moreover, BIC's are shown to exist near the ENZ condition of SiO$_2$, when these layers can act like ideal mirrors for the central air gap justifying the mechanism to be of Fabry-Perot type. Thus a continuous tuning of the operating point along one of the dispersion branches of the leaky modes eventually leads to the BIC. These results are validated by direct calculations of the reflectivity spectra, along with spatial eigen mode profiles depicting complete confinement. The results can be useful for having a deeper understanding of the BIC in such structures as the limiting case of the leaky modes for very rigid set of system parameters.
	\par
\vspace{0.5cm} 
\noindent
{\bf Acknowledgment} \\
One of the authors (SDG) would like to thank Rishav Banduri for help in calculations.

\bibliographystyle{ieeetr}
\bibliography{Combined}

\end{document}